An Essay on Interactive Investigations of the Zeeman Effect in the Interstellar Medium

Lauren Woolsey

Harvard University

60 Garden St, M.S. 10
Cambridge, MA 02138

lwoolsey@cfa.harvard.edu

Lauren Woolsey is a fourth-year graduate student in the astronomy department at Harvard. Her research interests focus primarily on solar wind modeling, but she enjoys working on interesting and informative side projects. She plans to seek a tenure-track teaching position at the college level upon graduating at the end of her fifth year.

TITLE: An Essay on Interactive Investigations of the Zeeman Effect in the Interstellar Medium

ABSTRACT: The paper presents an interactive module created through the Wolfram Demonstrations Project that visualizes the Zeeman effect for the small magnetic field strengths present in the interstellar medium. The paper provides an overview of spectral lines and a few examples of strong and weak Zeeman splitting before discussing the module in depth. Student discovery is aided with example situations to investigate using the interactive module, which is targeted at the upper undergraduate or early graduate level. This module (http://demonstrations.wolfram.com/TheZeemanEffectInTheInterstellarMedium), which uses free software, can be used in classroom activities or as a means of introducing students to the Wolfram Demonstrations Project as a learning resource.

INTRODUCTION

In electricity & magnetism courses, one of the more difficult concepts for students to visualize is magnetic field structure. Scientists can observe the direct effects of magnetic fields, but not the magnetic field itself. In order for students to have a better understanding of this topic, it is imperative that they see for themselves such direct effects (Dori, Hult, Breslow, & Belcher, 2007). While classroom demonstrations are incredibly useful, providing a larger context is also desirable. This can be provided by interactive simulations, which have been shown to increase student learning in physics (Weiman, Adams, & Perkins, 2008). This essay will describe one of these contexts and an interactive module designed for upper-level undergraduates and early-level graduate students to use for investigating the subject.

One of the largest contexts available in physics is application to astrophysics. There are a limited number of ways to measure magnetic field in the interstellar medium. The first uses linear polarization of starlight or scattered light from dust. This method provides a partial direction; like an arrow without an arrowhead, the vector is narrowed down to two possible directions. The downside of this method is this ambiguity in direction, because while many physical processes can create random field directions that may look aligned, magnetic forces can play a bigger role in regions of space where the field is coherent, pointing in the same direction over a large area.

Other methods that can probe the true direction of the magnetic field and determine its coherence are Faraday rotation and Zeeman splitting, the latter of which can be done in undergraduate labs (Blue, Bayram, & Marcum, 2010; Olson & Mayer, 2009). Faraday rotation can be used when there is a strongly linearly polarized background source and has many applications (Heiles, 1976; Oppermann, Junklewitz, & Robbers, 2012) The Zeeman effect can measure both direction and strength of the magnetic field, which is incredibly useful. However, for use in the interstellar medium (ISM), atoms or molecules with large magnetic moments need to be present in a high enough density for it to work. Therefore, as Bryan Gaensler once stated, "when you can use it, it's gold, but there are only very limited places in the universe where Zeeman splitting actually works" (colloquium at Harvard-Smithsonian Center for Astrophysics, 21-Mar-2013). This essay focuses on the use of the Zeeman effect in relatively dense portions of the ISM like molecular clouds.

OVERVIEW OF SPECTRAL LINES AND ZEEMAN SPLITTING

When spectral lines come up in astronomy and physics courses, instructors are most commonly discussing the absorption and emission lines created when an electron changes energy levels in an atom (Rybicki & Lightman, 2004). There are, however, subtler processes that also create spectral lines. A well-known example in astronomy is created by neutral Hydrogen (HI; Dickey & Lockman, 1990) When the spin of the lone electron in an HI atom flips from being parallel with the proton in the nucleus to being anti-parallel, a photon with a wavelength at 21 cm is emitted. The spin-flip transition of HI and other atoms or molecules are often observed in the radio due to the small energy level differences, and I will discuss why some of these long wavelength spectral lines are ideal for observing the Zeeman effect in the ISM.

These subtle processes that can produce long-wavelength spectral lines are often polarized. Components that are elliptically polarized will be important for magnetic field measurements of the ISM. The amount of splitting between two angular momentum quantum numbers for a single line is given by $\Delta \nu_Z = (g \mu_B B / h)$, where $\mu_B$ is the Bohr magneton, h is the Planck constant, and g is the Lande g-factor (Heiles, Goodman, McKee, & Zweibel, 1993).

The Sun is an example of the strong Zeeman effect, where splitting of the lines is strong enough to be seen in the overall intensity and the amount of splitting is directly measured from solar spectra using the equation above. This effect is strong enough to be measurable in undergraduate physics labs (Ratcliff, Noss, Dunham, Anthony, Cooley, & Alvarez, 1992). Such visible splitting of the field lines is rare in other astronomical observations, where field strengths tend to be much lower. Observations of the magnetic field strength at the Sun's photosphere, commonly referred to as magnetograms, are often used as boundary conditions for codes that model the 3D magnetic field of the Sun out to 1 AU. The accuracy of these boundary conditions is essential, as the Sun's magnetic field plays an important role in the heating of the corona and acceleration of the solar wind (Woolsey & Cranmer, 2014).

When the magnetic field is not strong enough to split the spectral lines completely, the process is referred to as the weak Zeeman effect. Heiles, Goodman, McKee, and Zweibel (1993) compile an extensive list of candidates for observations of the weak Zeeman effect, which I reproduce for only the most commonly used candidates in Table 1. Here, $b = (2 g \mu_B / h)$ Hz/μG such that the equation above can be rewritten as $\Delta v_Z = (b B / 2)$.

| Species | Transition | Line v (GHz) | b (Hz/μG) | $n_H$ (cm$^{-3}$) |
|---|---|---|---|---|
| HI | $^2\Sigma_{1/2}$, F=1-0 | 1.42 | 2.8 | Low (100-300) |
| OH | $^2\Pi_{3/2}$, F=1-1 | 1.665 | 3.27 | Clouds: $10^3$, masers: $10^7$ |
| OH | $^2\Pi_{3/2}$, F=2-2 | 1.667 | 1.96 | Clouds: $10^3$, masers: $10^7$ |
| H$_2$O | Hyperfine ($6_{16} - 5_{23}$) | 22.235 | 0.0029 | Masers: up to $10^9$ |

*Table 1: Common Species used for ISM Zeeman Observations*

In Table 1, the density $n_H$ is representative of the density of the environment in which each species is observable. Note that each of the lines listed, which are the best candidates for detecting Zeeman splitting, all have frequencies of 22 GHz or lower, which means very long wavelength observations (visible light has frequencies on the order of 500 THz).

The environments that can produce measurable Zeeman effects are denser than much of the ISM. The outer parts of molecular clouds are traced by excess 21-cm emission for HI and have densities of a few hundred molecules per cubic centimeter. OH is found in molecular clouds and is observable for densities up to 2500 cm$^{-3}$. Zeeman splitting in OH emission has also been observed in masers with strong magnetic fields. Hyperfine transition lines of H$_2$O are also seen in masers (Elitzur, 1992). In these extremely high density masers, the magnetic field can be strong enough to classically split the spectral line and the total magnetic field strength can be observed just as for solar magnetograms. However, the magnetic fields commonly observed in the interstellar medium in molecular clouds are low enough that the splitting of the spectral line is not great enough to produce a measurable effect. In this case, observers must turn to Stokes parameters. Right circular polarization and left circular polarization of spectral lines can be measured separately; the sum of these is called the Stokes I parameter and the difference is called the Stokes V parameter. With no magnetic field present, the Stokes V spectrum should be a flat line. However, if a magnetic field is present, the "V-spectrum" can be fit by the derivative of the "I-spectrum" scaled up by the strength of the magnetic field along the line-of-sight as follows: $V = (\Delta v_Z \, dI/dv)$ (Heiles, Goodman, McKee, & Zweibel, 1993; Sarma, Brogan, Bourke, Eftimova, & Troland, 2013). The interactive module I present in this paper plots these Stokes parameters as astronomers observe them (Woolsey, 2013).

**THE INTERACTIVE MODULE**

The module I have created to help students investigate the Zeeman effect has been written in Mathematica, converted to a Computational Document Format (CDF), and uses free software from Wolfram to run and use. This software, the Wolfram CDF Player can be downloaded at www.wolfram.com/cdf-player/. The module has several options that the user can change. They are seen as a pull-down menu and five slider bars in Figure 1:
1. Lines: the user can choose a spectral line from the list presented in Table 1.
2. Temperature: a hotter environment produces thermally broadened spectral lines.
3. Turbulent Velocity: turbulence adds non-thermal Doppler broadening to spectral lines.
4. Line-of-Sight Magnetic Field: this is what produces the Zeeman splitting.
5. Y-axis Bounds for Stokes V: the magnitude of the Zeeman effect ranges depending on inputs, so this allows the user to "zoom in" to a range where the effect is observable.
6. Signal-to-Noise Ratio: longer observations on a telescope produce a higher signal-to-noise, so the exposure time we need to see a clear signal for given inputs may differ, and this simulates the observational constraint.

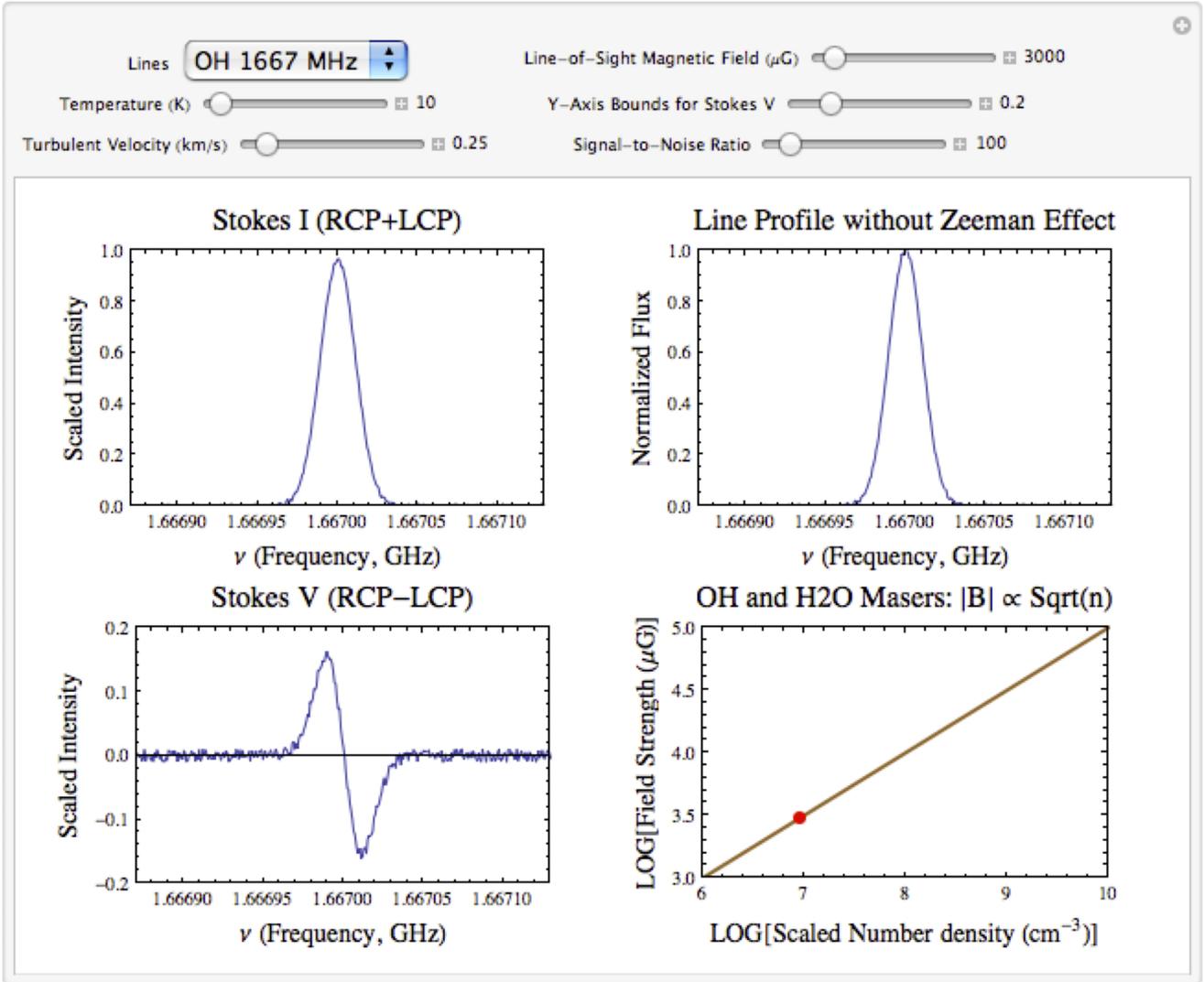

*Figure 1: A screenshot of the complete interactive module.*
The module plots Stokes I and Stokes V, discussed in the previous section. There is also a plot of the line profile without Zeeman effect, which is meant to provide a baseline for comparison with the Stokes I parameter. If there is no magnetic field, the top two plots (Stokes I and line profile without Zeeman effect) will be identical, and the Stokes V line will be flat. The flux is normalized to peak at a unitless value of 1 because we have not made any assumptions about the source brightness and distance or the telescope used. These details are beyond the scope of this module and are not necessary to understand the Zeeman effect in a general sense. The final plot represents a rough idea of how the magnetic field strength is related to the density of OH and $H_2O$ masers. Observations indicate a weak proportionality of the magnitude of line-of-sight magnetic field strength with the square root of $n_H$ (Crutcher, 2010). The relation arises from the effect of flux freeze-in, where magnetic field lines are dragged along by the dense medium above a threshold density.

This should serve as a good starting point to start playing with the module and discovering the difficulty of making different observations of the magnetic field in the interstellar medium. To provide additional guidance, Table 2 lists typical values of structures in the ISM. The turbulent velocity is approximated by $\sigma_v \sim 1.1$ (L [pc])$^{0.38}$ km/s (Larson, 1981).

| Region | Size, L (pc) | Density, n (cm$^{-3}$) | Temperature (K) | $\sigma_v$ (km/s) |
|---|---|---|---|---|
| Giant Molecular Cloud | 100 | 100 | 50 | 0.5 |
| Dark Cloud | 10 | 1000 | 20 | 0.3 |
| Core | 0.3 | 10000 | 10 | 0.2 |

*Table 2: Typical Properties of Structures in the ISM*

Students can use the module to experiment with similar values in the interactive module to get a sense of how difficult observations of the Zeeman Effect in the interstellar medium can be. The ISM has magnetic field strengths that typically range from 1 microgauss to tens of milligauss (0.1 nT to several hundred nT).

I have written a few example situations that students might investigate individually or during a short classroom investigation, but there are numerous uses of this module beyond these questions.

1. A typical Giant Molecular Cloud (GMC) might be 100 pc across, with an average temperature of 50 K (-370 degrees Fahrenheit!) and turbulent velocity of roughly 0.4 km/s (900 mph!). If line-of-sight magnetic fields are measured to be on the order of 200 microgauss by observing neutral hydrogen's 21-cm line, what is the ratio between peak intensity of Stokes V to Stokes I?
2. Astronomers set out to observe the OH 1665 MHz line in a maser. If the temperature is 300 K (typical room temperature on Earth, relatively hot for the interstellar medium) and there is no turbulence, at what magnetic field strength can we distinguish two completely separate spectral lines in the Stokes I spectrum? What if the turbulent velocity were 0.5 km/s? What if the turbulent velocity were as high as 1 km/s?
3. Let's see how difficult using the Water line at 22 GHz is for measuring the Zeeman effect. Play around with the five sliding controls until you have the clearest Stokes V signal. What were your values? Explain the effect of changing each slider on the signal you observe.

Students should take note of the relative intensity of the Stokes V to the Stokes I and the amount of signal-to-noise required for a clear signal in the Stokes V. Regions of the ISM where such observations are possible make up only a small fraction of the total volume. This module can therefore give students a sense of the difficulties faced in observations compared to the clear and straightforward theory they learn in classes. For advanced students, the module can also be used to reproduce scientific results by determining the magnetic field strength of a specific region based on the shape of a measured signal and the associated environmental properties.

## CONCLUSIONS

I present an interactive module that demonstrates the Zeeman effect, one of the few ways that astronomers can measure the strength and direction of magnetic fields in the interstellar medium. The module is an effective way for students to understand both the ways that magnetic field can be measured and the methods that scientists use to learn about the interstellar medium. It can easily augment classroom lectures and undergraduate labs in spectroscopy. The module was created using Mathematica and the powerful Computing Document Format (CDF) from Wolfram, which can be read with free software (Wolfram CDF Player). This module is available online (Woolsey, 2013). While I hope the specific module presented in the paper can find ongoing uses in astronomy courses, I also want this module to serve as an example of the capabilities of CDF. For students comfortable with Mathematica, the Wolfram Demonstrations Project is an incredible resource for learning modules and can be used for in-class projects, which was the origin of the module presented here.

## ACKNOWLEDGEMENTS

This material is based upon coursework under the advisement of Alyssa Goodman and supported by the National Science Foundation Graduate Research Fellowship Grant No. DGE-1144152. The author thanks Phil Sadler and the anonymous reviewer for helpful comments and suggestions.